\documentstyle[11pt,newpasp,twoside,epsf]{article}
\def\ts     {\thinspace}
\def\kms    {\ifmmode{{\rm \ts km\ts s}^{-1}}\else{\ts km\ts s$^{-1}$}\fi}
\def\mo     {M$_{\odot}$}
\def\hi     {H{\small I}}

\def\edcomment#1{\iffalse\marginpar{\raggedright\sl#1\/}\else\relax\fi}
\reversemarginpar

\begin{document}
\title{Atomic and Molecular Gas near NGC\,3077: The Making of a New Dwarf Galaxy?}
 \author{Fabian Walter}
\affil{California Institute of Technology,
  Astronomy Department 105-24, Pasadena, CA 91125, USA}
\author{Andreas Heithausen}
\affil{Radioastronomisches Institut der Universit\"at Bonn, Auf dem H\"ugel 71,
53121 Bonn, Germany}

\begin{abstract}
Using the IRAM 30\,m radio telescope we have mapped the tidal arm
feature south--east of NGC\,3077 where we recently detected molecular
gas in the CO ($J$=1$\to$0) and (2$\to$1) transitions. We find that
the molecular gas is much more extended than previously thought
(several kpc).  The CO emission can be separated into at least 3
distinct complexes with equivalent radii between 250\,pc and 700\,pc
-- the newly detected complexes therefore range among the largest
molecular complexes in the local universe. Mass estimates based on
virialization and employing an $X_{\rm CO}$ factor yield a total mass
for the complexes of order $4\times10^7$\,\mo, i.e. more than the
estimated molecular mass within NGC\,3077 itself. This implies that
interactions between galaxies can efficiently remove heavy elements
and molecules from a galaxy and enrich the intergalactic medium.  A
comparison of the distribution of \hi\ and CO shows no clear
correlation. However, CO is only found in regions where the \hi\
column density exceeds $1.1\times10^{21}$\,cm$^{-2}$. \hi\ masses for
the molecular complexes mapped are of the same order as the
corresponding molecular masses.  Since the complexes have all the
ingredients to form stars in the future, we are thus presumably
witnessing the birth of a dwarf galaxy. This process may have
dominated the creation of dwarf galaxies at larger look--back times.
\end{abstract}
\section{Introduction}

Over the last decades it has become clear that tidal forces during
close encounters of galaxies can redistribute large masses as long
tails or bridges between the interacting galaxies.  The idea that part
of these newly formed structures could form self--gravitating entities
was proposed by Zwicky (1956). Numerical simulations
(e.g. Elmegreen et al.\ 1993; Barnes \& Hernquist
1996) support this scenario.

Observationally, tidal systems are best traced by the neutral gas
phase (by means of \hi\ observations) since it is this extended
component of a galaxy which is most easily disrupted by interactions.
Also, much work has been done in studying the tidal arms of
interacting galaxies at optical wavelengths to find regions of active
star-formation (`tidal dwarfs').  Impressive examples of tidal arms
with on-going star-formation are, e.g., the Antenna galaxy
(NGC\,4038/39, Mirabel et al. 1992) and the Superantenna system
(Mirabel et al.\ 1991).  Based on an optical study of 42 Hickson
compact groups of galaxies Hunsberger et al.\ 1996) speculate that up
to 50\% of the dwarf galaxies in such compact groups might be created
by tidal interaction among giant parent galaxies.

\begin{figure}[t] %[htb]
  \epsfxsize=16cm
    \begin{center}
        \leavevmode \epsfbox{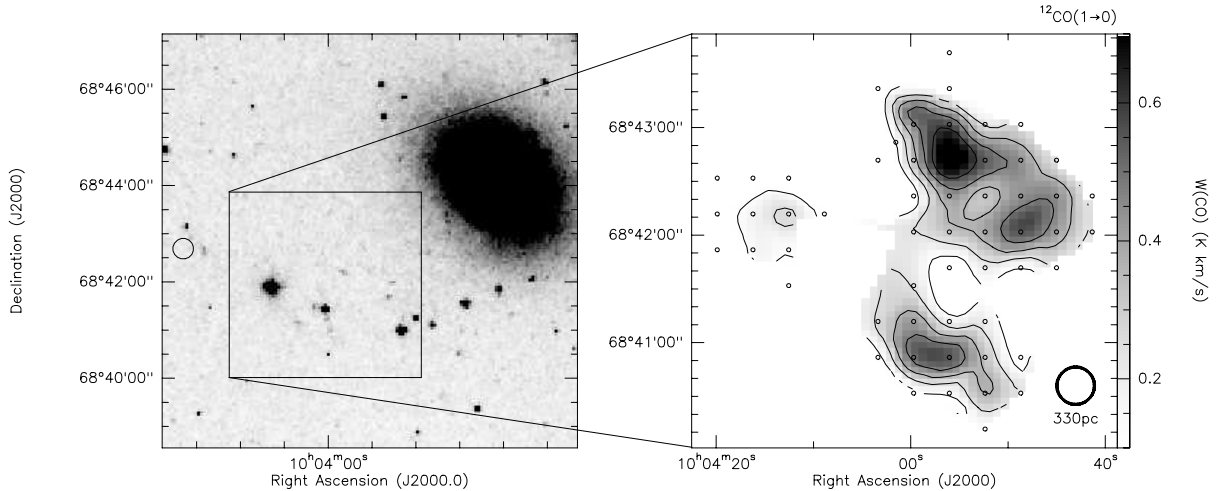}  
    \end{center}
    \vspace{-0.5cm}
\caption{Overview of the observed region. {\em Left:} optical image of
NGC\,3077 (elliptical object to the north--west) and its surrounding
(taken from the digitized sky survey). The box indicates the region
which we mapped in CO. {\em Right:} Blowup of our integrated CO map
(J=1$\to$0 transition). Contours are every 0.12\,K\,\kms\ ($2\sigma$)
starting at 0.12\, K\,\kms. Observed positions are indicated as small
circles. The beamsize is indicated in the lower right corner of the
map; it corresponds to 330\,pc at an adopted distance of
3.2\,Mpc.\label{comap}}
\end{figure}

Surprisingly little is known about molecular gas in tidal arms around
interacting galaxies.  However, this is an important issue since
molecular clouds are the places where stars are born. The distribution
of molecular gas in quiescent extragalactic objects (such as tidal
arms) therefore gives clues as where to expect star formation to
commence in the future.

The most extended molecular complex in tidal arms of interacting
galaxies was recently discovered by us near NGC\,3077, member of the
M\,81 triplet (see van der Hulst 1979, Yun 1994, Walter \& Heithausen
1999 for HI mapping of this tidal arm). This molecular complex is of
particular interest since, although it is huge, hardly any star
formation seems to be associate with it.

In the following we present a detailed follow-up study of this complex
(Heithausen \& Walter 2000). 

\section{Observations \& Results \label{observations}}

The CO ($J$=1$\to$0) and (2$\to$1) transitions have been observed in
July 1999 using the IRAM 30\,m radio telescope. The beam sizes are
22$''$ at 115\,GHz and 11$''$ at 230\,GHz (corresponding to 330\,pc
and 165\,pc at the adopted distance of NGC\,3077, 3.2\,Mpc). The
mainbeam efficiencies are $\eta_{\rm mb}(115 \rm GHz)=0.84$ and
$\eta_{\rm mb}(230 \rm GHz)=0.55$, respectively.  Pointing accuracy
was better than 5$''$. Spectra were obtained with a velocity
resolution of 0.8\,\kms\ at 115\,GHz and 0.4\,\kms\ using
autocorrelator spectrometers. In total, we observed 47 individual
positions simultaneously in both transitions with a wobbling secondary
mirror; wobbler throw was $\pm4'$ in azimuth.  The spacing between
individual positions is 20$''$ ($\sim$ the size of the beam at
115\,GHz).

Fig.~1 gives an overview over our mapping results. An optical image is
shown on the left (as obtained from the Digitized Sky Survey, DSS).
Our CO detections (as indicated by the box and the `X') are clearly
located outside NGC\,3077 (the elliptical object north--west of the
centre). The CO emission is clearly extended over several kpc
(1$'\sim$1\,kpc) and can be subdivided into at least two separate
complexes. A further cloud (`X') has been detected outside the mapped
area -- this area has only partly been mapped by us as yet.

\begin{figure}[t] %[htb]
  \epsfxsize=9cm
    \begin{center}
        \leavevmode \epsfbox{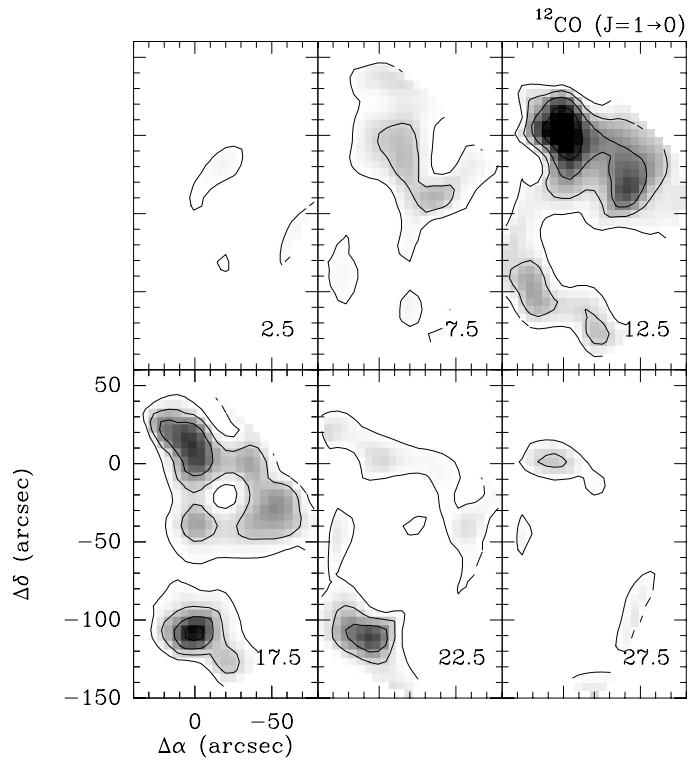}  
    \end{center}
    \vspace{-0.5cm}
\caption{
Channel maps of complexes \#1 and 2, averaged over 5\,\kms. Center
velocities of each channel are indicated in the lower right
corner. Contours are every 0.012 K ($2\sigma$) starting at
0.012\,K. Offsets are relative to $\alpha_{J2000}=10^h03^m56.\!^s0$;
$\delta_{J2000}=68^\circ42' 41.\!''8$ }
\label{cochan}
\end{figure}

The velocity structure of complexes \#1 and \#2 is visible in the
channel maps displayed in Fig.~2. Cloud \#1 is extended
from the south--west to north--east. Complex \#2 is the southern more
compact object. It shows evidence for further substructure.  We expect
these complexes to break up in more little clouds when observed at
higher spatial resolution. Throughout the paper we discuss only properties of
the whole complexes and do not subdivide them further, although
especially for complex \#1 there is evidence for substructure in both
the channel maps (Fig.~2) and in the integrated map
(Fig.~1).

\section{Physical parameters of the molecular gas \label{physpar}}

The molecular complex near NGC\,3077 discussed here is larger than
many complexes in other galaxies; e.g., the complex within NGC\,3077
itself has a size (FWHM) of only 320\,pc (Becker et al.\ 1989).  Cohen
et al.\ (1988) report a full extent of the 30 Dor complex in the LMC
of 2400\,pc which is believed to be one of the largest CO complexes in
the local universe.  Most Galactic molecular clouds have sizes below
60\,pc (Solomon et al. 1987). The Orion A \& B complexes (see Dame et
al. 1987) placed at the distance of NGC\,3077 would be detectable with
the sensitivity of our observations, however just in one single
spectrum.  {\it The complex near NGC\,3077 thus belongs to one of the
largest concentrations of molecular gas in the local universe.}

\subsection{Molecular cloud masses \label{comass}}

One important yet difficult to determine physical parameter is the
mass of a molecular complex. Determination of the mass is usually
based on either the assumption of virialization and/or application of
a $X_{\rm CO}=N({\rm H}_2)/W_{\rm CO}$ conversion factor. In this
section we apply both methods and discuss their pros and cons.

To estimate the $X_{\rm CO}$ factor we compare the CO luminosity for a
given line width with that of a cloud with the same line width but
with known molecular mass (e.g. Cohen et al.\ 1988). Using Fig.~2 of
Cohen et al.\ we find that complex \#1 lies within the range of values
span by Galactic clouds, whereas complex \#2 lies in the range span by
LMC clouds. This indicates that the tidal arm clouds have CO
luminosities for a given line width in between those of Milky Way and
LMC clouds. The $X_{\rm CO}$ factor of the Milky Way
($\sim2.5\times10^{20}$\,cm$^{-2}$\,(K\,\kms)$^{-1}$) is a factor of
6.7 lower than that of the LMC
($1.7\times10^{21}$\,cm$^{-2}$\,(K\,\kms)$^{-1}$, Cohen et al.\
1988).  Thus we use an $X_{\rm CO}$ value for the tidal
arm clouds which is in between both values, $X_{\rm
CO}=8\times10^{20}$\,cm$^{-2}$\,(K\,\kms)$^{-1}$.  This leads to
molecular masses for complex \#1 of $M_{X,1}=1.7\times10^7$\,\mo\ and
for complex \#2 of $M_{X,2}=0.4\times10^7$\,\mo, (corrected for the
contribution of He).

If we adopt a $1/r$ density profile through the clouds the assumption
of virialization leads to masses for complex \#1 of $M_{\rm
vir,1}=2.4\times10^7$\,\mo\ and for complex \#2 of $M_{\rm
vir,2}=1.6\times10^7$\,\mo.  Given the uncertainties in both methods
the masses for complex \#1 agree well, however those for complex \#2
are discrepant by a factor of 4.  At this point we can only speculate
which method gives the better estimate for the true molecular masses.

\subsection{The relation to the \hi\ gas \label{HI}}

\begin{figure}[t] %[htb]
  \epsfxsize=9cm
    \begin{center}
        \leavevmode \epsfbox{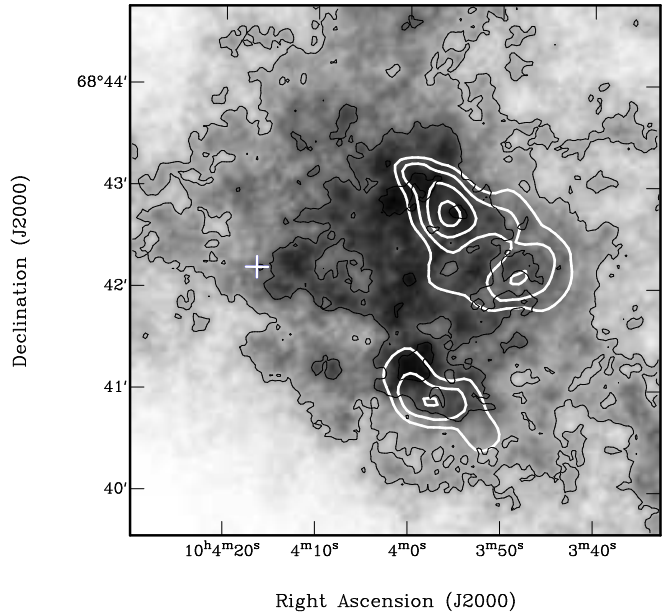}  
    \end{center}
    \vspace{-0.5cm}
\caption{Overlay of 
CO ($J=1\to0$) contours on an integrated \hi\ map (grey scale).  Thick
white (CO) contours are every 0.14\,K\,\kms\ starting at 0.28\,K\kms.
Thin black contours represent \hi\ column densities of 1$\times$,
1.5$\times$, and 2$\times10^{21}$\,cm$^{-2}$. The + marks the position
of complex \#3.\label{hi-co}}
\end{figure}

Fig.~3 shows a comparison of the distribution of the CO gas
with that of atomic hydrogen. We use our high-angular resolution
(13$''$) 21\,cm map as obtained with the VLA (Walter \& Heithausen
1999). It is obvious, that there is no direct correlation between the
intensities of \hi\ and CO. CO is not concentrated to the maximum peak
of the \hi\ column density, but rather found on the outer area of the
$1.5\times10^{21}$\,cm$^{-2}$ contour.  The average \hi\ column
density associated with complex \#1 and 2 is
$1.6\times10^{21}$\,cm$^{-2}$, that for complex \#3 is
$1.4\times10^{21}$\,cm$^{-2}$. 

The threshold where to find molecular gas depends on both metallicity
and radiation field (e.g. Pak et al. 1998). The shielding
\hi\ column density in the tidal arms around NGC\,3077 is slightly
higher than values in other galaxies. In the Milky Way the H$_2$
threshold appears at $0.6\times10^{21}$\,cm$^{-2}$ (Savage et
al. 1977). CO observations of M\,31 suggest a threshold of about
10$^{21}$\,cm$^{-2}$ (Lada et al. 1988). Young \& Lo 1997) found a
threshold of $0.1-0.2\times10^{21}$\,cm$^{-2}$ for the dwarf
elliptical galaxies NGC\,185 and NGC\,205; they attribute the
difference to the Milky Way value and M\,31 to the lower radiation
field in these dwarf ellipticals. We only can speculate what causes
the higher threshold in our complex. The radiation field is probably
low because we do not see strong associated star forming regions.

The total \hi\ mass of the tidal arm feature around NGC\,3077 was
found to be $M$(\hi)=$(3-5)\times10^8$\,\mo\ (van der Hulst 1979,
Walter \& Heithausen 1999), depending on integration borders. If we
regard only the areas where CO is detected we find an \hi\ mass for
complex \#1 of about $2.7\times10^7$\,\mo\ and for complex \#2 of
about $0.8\times10^7$\,\mo. These atomic masses are comparable to the
molecular masses derived by our adopted $X_{\rm CO}$ factor (see
Sec.~\ref{comass}).

\section{Conclusions \label{conclusions}}

Our new observations have revealed extensive molecular gas in the
tidal arms near NGC\,3077. The CO emission is much more extended than
previously thought -- the detected complexes range among the most
extended complexes in the local universe. We have detected at least
three independent complexes of molecular gas. The complexes are most
probably gravitationally bound objects and have formed {\it in
situ}. For the largest of the complexes our multi-transition CO study
shows that the gas must be 10\,K cold, thus probably not much star
formation is going on there.  Whether or not the chain of blue stars
(the `Garland') found in this region (Karachentsev et al.\ 1985a,
1985b) is associated with the molecular complex is an open question
which is currently under investigation by us.

The fact that CO is found in tidal arms implies that galaxy
interactions can efficiently remove enriched material from a galaxy's
body hence influencing it's chemical history. This also has important
implications for the chemical enrichment of the intergalactic medium
(IGM), especially at larger look--back times in the universe where
galaxy interactions may have been more frequent.

Since the newly discovered complexes have all the ingredients to form
stars in the future (i.e., atomic and molecular gas), our new
observation confirm our previous speculation (Walter \& Heithausen
1999) that we are witnessing the birth of a dwarf galaxy where star
formation might start in the near future. We are therefore in the
fortunate situation to witness a process which may have created a
substantial number of today's dwarf galaxies in the past.

\end{document}